   \documentclass[journal]{IEEEtran}
\usepackage{amsfonts,amsmath,amssymb}
\usepackage{relsize}
\usepackage{cite}
\usepackage{graphicx}
\usepackage{url}
\usepackage{epstopdf}
\usepackage{bm}
\usepackage{xcolor}
\usepackage{arydshln}
\usepackage{algorithm}
\usepackage{algorithmicx}
\usepackage{algpseudocode}
\usepackage{setspace}
  \newtheorem{remark}{Remark}

\begin{document}

\title{Pooling is not Favorable: Decentralize Mining Power of PoW Blockchain Using Age-of-Work}

\author{\authorblockN{Long Shi, Taotao Wang, Jun Li, and Shengli Zhang}

% \authorblockA{$^\star$Science and Math Cluster, Singapore University of Technology and Design, Singapore\\
%  $^\dag$School of Computing \& Communications, University of Technology Sydney, Australia\\
%  $^\ddag$College of Information Engineering, Shenzhen University, Shenzhen, China \vspace{-0.2in}

\thanks{L. Shi and J. Li are with the School of Electronic and Optical Engineering, Nanjing University of Science and Technology, Nanjing 210094, China (e-mail: slong1007@gmail.com; jun.li@njust.edu.cn).}

\thanks{T. Wang, and S. Zhang are with the College of Electronics and Information Engineering, Shenzhen University, Shenzhen 518060, China (e-mail: ttwang@szu.edu.cn; zsl@szu.edu.cn)}

\thanks{L. Shi and T. Wang contributed equally to this work.}

%  }
 }

\maketitle

\setcounter{page}{1}
\begin{abstract}
As the underlying consensus protocol of Bitcoin and Ethereum blockchain, Proof-of-Work (PoW) features a cryptographic mathematical puzzle whose solution is easy to verify but extremely hard to solve. Under PoW, miners maintain the security of blockchain by devoting computing powers to solve the puzzle; the miner who has solved the puzzle successfully generates a block, along with a reward (e.g., a set of cryptocurrency). The average waiting time to generate a block is inversely proportional to the computing power of the miner. To reduce the average block generation time, a group of individual miners can form a mining pool to aggregate their computing power to solve the puzzle together and share the reward contained in the block. However, if the aggregated computing power of the pool forms a substantial portion of the total computing power in the network, the pooled mining undermines the core spirit of blockchain, i.e., the decentralization, and harms its security. To discourage the pooled mining, we develop a new consensus protocol called Proof-of-Age (PoA) that builds upon the native PoW protocol. The core idea of PoA lies in using Age-of-Work (AoW) to measure the effective mining period that the miner has devoted to maintaining the security of blockchain. Unlike in the native PoW protocol, in our PoA protocol, miners benefit from its effective mining period even if they have not successfully mined a block. We first employ a continuous time Markov chain (CTMC) to model the block generation process of the PoA based blockchain. Based on this CTMC model, we then analyze the block generation rates of the mining pool and solo miner respectively. Our analytical results verify that under PoA, the block generation rates of miners in the mining pool are reduced compared to that of solo miners, thereby disincentivizing the pooled mining. Finally, we simulate the mining process in the PoA blockchain to demonstrate the consistency of the analytical results.
\end{abstract}

%Our analytical results verify that the block generation rate of the mining pool in PoA is not proportional to its aggregated computing power and can even be much less than its proportion, thereby disincentivizing the pooled mining. Finally, we simulate the mining process in the PoA-enabled blockchain to demonstrate the consistency of the analytical results.


%\IEEEpeerreviewmaketitle

\section{Introduction}\label{sec:intro}

Blockchain was born as a supporting technology for secure, effective, and transparent digital asset transactions between peers \cite{nakamoto2008bitcoin}. Thanks to its ability to enable Byzantine agreement over a permissionless decentralized network, Blockchain has become the foundation of various applications that demand critical data security and integrity, such as FinTech, Internet of Things (IoT), and supply chains \cite{fanning2016blockchain, dai2019blockchain,abeyratne2016blockchain}.

Blockchain features three key components, i.e., transactions, blocks, and chain, where transactions are recorded in blocks that are linked together to form a chain. The goal of blockchain is to enable a public, shared, and distributed ledger among different trustless nodes. Towards this goal, various consensus mechanisms have been implemented to realize the global agreement \cite{natoli2019deconstructing}, which are paramount to guarantee the security and sustainability of the network. The most prevailing consensus mechanism is Proof-of-Work (PoW) \cite{nakamoto2008bitcoin}, which has been widely deployed in many dominating blockchain networks, such as Bitcoin and Ethereum. The native PoW protocol features a computational complex cryptographic math puzzle whose solution is easy to verify but extremely hard to solve. The nodes in the blockchain network exhaust massive computing resources to generate a solution (called nonce). The process is called mining, and those who take part in it are known as miners. Under PoW, the miner generates a block if it can solve the PoW puzzle with a valid nonce, and immediately broadcasts this block that contains the transactions and the nonce to the entire network. Each node updates its local blockchain by appending the newly mined block only if the block is validated by the majority of nodes in the network. In this way, the miner successfully mines a block.

In the blockchain network, the miner that successfully mines a block is rewarded with a set of cryptocurrency. Driven by this incentive mechanism, the miners contribute their computing powers to the mining process to harvest the reward. In this context, it is more likely for the miners with large computing power to gain the rewards by performing a large number of hash trials within a certain period. Consequently, it may take extremely long time for the solo miners with relatively small computing power to successfully generate a block. For example, in the current Bitcoin network, the chance for a solo miner even deployed with dedicated ASIC mining machine to mine a block is very tiny. As a result, this incentivizes a group of solo miners to aggregate their computing powers into a mining pool, where all participating miners mines towards the same target and share the reward. With this approach of pooled mining, the miners in the pool can greatly reduce their average waiting time to gain the rewards. While pooled mining can indeed bring significant benefits to the miners, the pooling of computing power undermines the core sprit of blockchain, i.e., decentralization{\footnote{The computing powers of the largest nine mining pools in the current Bitcoin blockchain account for over 90\% of the whole blockchain \cite{BTCcom}.}}, and harms its security. For example, a mining pool that occupies more than 50\% of the total network computing power can arbitrarily tamper with the data recorded on blockchain; a mining pool that occupies more than 25\% of the total network computing power can use the selfish mining strategy proposed in \cite{eyal2014majority} to increase its revenue to more than that of the honest mining.

To overcome this problem, we develop a new consensus protocol called Proof-of-Age (PoA) that builds upon the native PoW protocol. The objective of PoA is to disincentivize  centralized pooled mining so that the miners with small computing powers can participant in and earn rewards from the mining competition. The core idea of PoA lies in using Age of Work (AoW) to measure  effective mining period that the miner has devoted to maintaining the security of blockchain. Unlike in the original PoW protocol, in our PoA protocol, the miner benefits from its effective mining period even if it has not successfully mined a block that contains a valid nonce. In particular, PoA exploits the accumulated AoW to reduce the mining difficulty if the miner can prove its effective work for a certain period without gaining any reward. Specifically, if a miner really works but it fails to mine a block, its AoW still increases; and its mining difficulty will be reduced if its accumulated AoW reaches a predesigned threshold; the AoW of a miner reduces to the smallest immediately it successfully generates a block, resulting in the maximum mining difficulty in the next mining round\footnote{The idea of PoA originates from the random access problem of wireless channels, where transmitters access the channel to transmit their information in a distributed manner according to some random access protocol. In \cite{chen2020age}, the concept of Age-of-Information (AoI) \cite{kosta2017age} is used to design the random access protocol. If a transmitter successfully access the wireless channel by sending an information, its instantaneous AoI returns to one; otherwise, its instantaneous AoI is increased by one. AoI is a metric to evaluate the freshness of the information hold at the transmitter for transmissions. The wireless network can use AoI to coordinate the random access to achieve some kind of fairness among transmitters \cite{chen2020age}.

In PoA based blockchain, we treat blocks as information transmitted by miners. Then, we can use a metric similar to AoI (i.e., AoW in PoA) to adjust the mining difficulties of miners. The situation is very similar to the random access of wireless channels. If a miner has successfully mined a block and added the block onto the blockchain, it corresponds to one-time successful access of a transmitter to wireless channel. The intuition of our design is that the miners with relatively lower AoW should access the channel (i.e., add blocks onto the blockchain) with smaller probability, while it is easier for the miners with higher AoW to mine a block.
}. The main contributions of this paper are listed as follows:

\begin{itemize}
	
	\item \emph{Protocol design:} We put forth the design of PoA protocol including the hash puzzle, data structure, mining targets, and difficulty adjustment. Different from native PoA, we introduce a new parameter called age ring into the input of hash puzzle. From the data structure of age ring, PoA also enables a chain of age ring, since the generation of age ring in the current mining round depends on that in the previous round. In addition, we set two different targets for the mining of age ring and block  respectively, such that the generation of an age ring is much easier than that of a block. For each mining round, the AoW of a miner will be increased if the solution of hash puzzle falls into the predesigned target region (i.e., the associated age ring in this round is effective), while the AoW of the miner will be recounted from the initial value immediately   the miner successfully generates a block. Furthermore, we propose a difficulty adjustment strategy of block mining, where the miner will be rewarded by a smaller  mining difficulty if its accumulated AoW reaches a predesigned threshold.

	\item \emph{Model analysis}:  We model the block generation process of PoA based blockchain by using a continuous time Markov chain (CTMC) model. In the proposed model, we decouple the highly correlated mining process of all miners as two parties, i.e., a typical miner and treat the impact of other miners as a whole from the typical miner's perspective. Meanwhile, we approximate the block generation process of other miners as a single exponential distribution with a rate parameter. In this context, the CTMC model is well-suited to capture the mining process of PoA. Furthermore, based on the CTMC model, we analyse the block generation rates of the mining pool and each solo miner respectively. In particular, our analytical results verify that the block generation rate of the mining pool in PoA is much less than that in PoW, thereby deincentivizing the pooled mining under PoA. It is shown that the block generation rate of the mining pool under PoA can be reduced to 75\% of that under PoW by a proper difficulty adjustment strategy. Next, we investigate the distributions of block inter-arrival times of the mining pool and solo miners respectively. Particulary, we verify that the block inter-arrival time of all solo miners is exponentially distributed.
	
	\item  \emph{Event-driven simulations}:  We simulate the mining process in the PoA based blockchain by using the event-driven method. To investigate the block generation rate, we generate the numerical results of the mining pool and each solo miner by the fixed point iteration method and compare the numerical results with the simulation results, and we show that the numerical results are consistent with the simulation results. Moreover, we show that the simulated distribution of the block generation times of all solo miners matches well with the theoretical exponential distribution, and the simulated distribution of the block generation times of the pool fits well with the analytical results. We finally compare the block generation rates of PoW and PoA in the simulations. The simulation results highlight that not only the block generation rate of the pool in PoA is much less than that in PoW, but also the variability of block inter-arrival times of the pool in PoA is much less significant than that in PoW.

\end{itemize}

The rest of this paper is organized as follows. Section II reviews the related work. Section III presents background on the PoW blockchain. Section IV introduces the PoA protocol. Section V proposes the CTMC model of PoA and analyzes the block generation rates of PoA blockchain. Section VI provides the numerical results. Finally, Section IV concludes this work.

\section{Related Works}\label{sec:ref}

A variety of consensus protocols have been developed to cope with the shortcomings of native PoW. One shortcoming of PoW is the low on-chain transaction processing capability. To increase the transaction processing capability of PoW blockchain, \cite{lewenberg2015inclusive, li2018scaling, sompolinsky2015secure, sompolinsky2016spectre, sompolinsky2018phantom} modify the data structure of the blockchain, and \cite{eyal2016bitcoin,rizun2016subchains, TierNolan, bagaria2019prism} decouple the functionality of the block. However, the focus of our PoA is not on improving the transaction processing capability. The aim of PoA is to deincentivize the centralization of computing powers caused by the pooled mining in blockchain.

To decentralize the pooled mining, the consensus protocols of Fruitchain \cite{pass2017fruitchains} decouples the transaction-carrying functionality of the blockchain. Fruitchain enables two independent mining processes on top of each other. In addition to the block mining, Fruitchain requires an additional PoW to mine a new type of block called ``fruit". More specifically, the fruit needs to point to an earlier block  not too far from the block that records the fruit. In this way, each block confirms a set of its recent fruits, while the fruits confirm transactions in this block. The fruit mining also requires solving partial PoW that corresponds to finding a nonce with a difficulty level much smaller than the block mining. Consequently, the solo miner with small computing power can reap from mining the fruits, once the fruits are confirmed by the block. Different from Fruitchain, PoA largely follows the data structure and communication protocol of native PoW with the cost of negligible overhead caused by age rings. Similar to the fruit mining in Fruitchain, the mining difficulty of an age ring is much smaller than that of a block. However, the age rings do not contain any transactions, but are used to validate the miner's AoW. Generally, there are three notable disadvantages of Fruitchain compared with our PoA. First, broadcasting the fruits in Furitchain increases the communication overhead of the network. Second, Furitchain needs to modify the consensus mechanism in the layer one of the blockchain, thus it has difficulty in  compatibility with the current PoW based blockchains. Finally, the rewards contained in the block need to be allocated to the fruits, resulting in instable profits earned by miners.

%The principle of functionality decoupling has been also adopted in BitcoinNG \cite{eyal2016bitcoin}. Bitcoin NG decouples transaction proposal and leader election, by electing a single leader to propose many transaction blocks till the next leader is elected by PoW. Beyond the transaction-carrying functionality decoupling, Prism proposes a total decoupling of transaction proposing, validation, and confirmation functionalities in the blockchain \cite{bagaria2019prism}. Compared with the functionality decoupled protocols, PoA resembles native PoW with smallest modification, which is more compatible with the current PoW blockchain.

Nowadays, two decentralized mining pools are proposed for Bitcoin and Ethereum blockchains. For Bitcoin, P2Pool \cite{P2pool} creates a new blockchain wherein the block is generated to record the shares of the miners in the pool. However, the P2Pool network needs to check a massive number of shares submitted from different participants in the pool in each mining round. Differently, the miners under the proposed PoA protocol are required to verify the age rings only from a single miner that are included in the newly generated block. For Ethereum, Smartpool in \cite{luu2017smartpool} runs a decentralized mining pool protocol as a smart contract. In spite of lower cost than centralized pools, miners still need to acquire Ether to pay for the gas when interacting with the smart contract in smartpool.

As an extension of PoW, Proof-of-Contribution (PoC) adjusts each miner's mining difficulty by introducing success times (i.e., the number of times the miner has added a valid block to the chain) \cite{xue2018proof}. The higher success times the miner accumulates, the smaller the mining difficulty of the miner becomes. Therefore, PoC motivates honest mining behaviour by rewarding the difficulty reduction. However, PoC cannot decentralize the pooled mining if the pool's work is legitimate and honest. Differently, the proposed PoA protocol rewards the partial PoW of the miner by adjusting its mining difficulty if its accumulated AoW meets a predesigned threshold.

To analyze the mining behaviors and the mining performances in blockchain networks, discrete time Markov chain (DTMC) was widely applied in the existing works. For example, \cite{eyal2014majority} employed DTMC models to analyze selfish mining strategies under PoW, i.e., to calculate a selfish miner's benefit as well as to determine the mining power threshold at which this attacker is beneficial.To extend the selfish mining, DTMC is also used in \cite{nayak2016stubborn} to study the mining revenue of the selfish miner under stubborn mining strategies. Later on, \cite{carlsten2016impact} modifies DTMC of selfish mining strategies by incorporating the impact of transaction fees and studies the corresponding mining performance. Moreover, \cite{sapirshtein2016optimal, sompolinsky2016bitcoin, gervais2016security} exploit the model of Markov Decision Process (MDP) to generalize selfish mining behaviors and find the optimal selfish mining behavior. However, as the number of miners in the network increases, using DTMC/MDP results in an unaffordable model complexity due to the huge state space and correlation among these miners. To address this issue, we employ the CTMC to model the mining process of PoA with a decoupling hypothesis. With the aid of the CTMC model, we can further derive the block generation rate and analyze the distribution of block inter-arrival time.

\section{Background of Blockchain and PoW}\label{sec:back}
This section presents the background of the PoW based blockchain.

\subsection{Data Structure of Blockchain}
% Blockchain is first proposed as the decentralized append-only public ledger for the first-ever decentralized cryptocurrency, Bitcoin.

Blockchain is a chain of blocks with each block containing a header and a set of transactions issued by the payers. Specifically, the issued transactions are broadcast over the blockchain network. Participants then collect and include these transactions into blocks and append them to the blockchain. The header of the block encapsulates the hash of the preceding block, the hash of this block, the merkle root of all transactions contained in this block, and a number called nonce that is generated according to the consensus protocol of PoW. Since each block must refer to its preceding block by placing the hash of its preceding block in its header, the sequence of blocks then forms a chain arranged in a chronological order. Thanks to the cryptographic chain structure, blockchain is tamper-resistant to modification of its past recorded data such that participants can only append new data to the tail-end of the chain of blocks. Fig. 1 illustrates the data structure of Bitcoin blockchain.
 \begin{figure}[t]
  \centering
        \includegraphics[height=0.55\columnwidth]{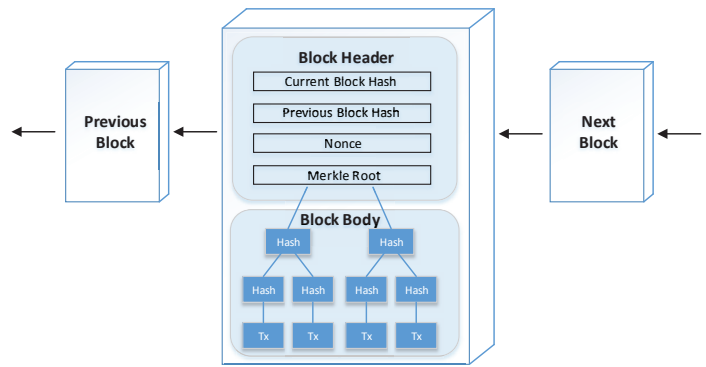}
       \caption{The data structure of the Bitcoin blockchain.}
        \label{fig:bitcoin}
\end{figure}

\subsection{Preliminaries of PoW}\label{sec:pro}

Bitcoin blockchain adopts the PoW consensus protocol among a number of participants to validate new blocks in a decentralized manner. In each round, the PoW protocol selects a leader to pack transactions into a block and appends this block to blockchain. To prevent adversaries from tampering blockchain, the leader selection must be approximately random. Since anonymity is inherently designed as a goal of permissionless blockchain, it must not be vulnerable to the sybil attack where the adversary simply creates many participants with different identities to increase its probability of being selected as the leader. To address the above issue, the key idea behind PoW is that a participant will be selected as the leader of each round with probability in proportion to its computing power.

In particular, blockchain implements PoW by using computational hash puzzles. To create a block at height $k$, the nonce placed into the header of the block must be a solution to the hash puzzle expressed by the following inequality \cite{wang2019survey}:
\begin{align}\label{eqn:hash}
h_k=H(b_{k-1},x_k,y_k)\leq T(D),
\end{align}
where $h_k$ is the hash of the candidate block at height $k$ and it is a bit stream of length $L$, $H(\cdot)$ is the cryptographic hash function, $b_{k-1}$ is the hash of previous block header at height $k-1$, $x_k$ denotes the solution string of the nonce, $y_k$ denotes the binary string assembled based on the other data contained in the candidate block (e.g., the Merkle root of the included transactions), $T(D)=2^{L-D}$ is a target value, and $D$ is the current difficulty level of the hash puzzle (i.e., the number of leading zeros in the hash of a valid candidate block). Using the blockchain terminology, the process of computing hashes to find a nonce is called mining, and the participants involved are called miners.

With a difficult level $D$  and the corresponding target $T(D)$, each single query to the PoW puzzle expressed in \eqref{eqn:hash} is an i.i.d. Bernoulli test whose success probability is given by
\begin{align}\label{eqn:suc}
Pr(x_k: H(b_{k-1},x_k,y_k)\leq T(D))=\frac{T(D)}{2^L}=2^{-D},
\end{align}
When $D$ is very large, the above success probability of a single query is very tiny. Moreover, it is known that, with a secure hash algorithm (e.g., the SHA-256 hash used for Bitcoin), the only way to solve \eqref{eqn:hash} is to exhaustively search every possible nonce. Therefore, the probability of finding such nonce to solve \eqref{eqn:hash} is proportional to the computing power of the miner. That is, the faster the hash function in \eqref{eqn:hash} can be computed in each trial, the more nonces can be tried per unit time.

Let $p=2^{-D}$ in \eqref{eqn:suc}. The probability that the miner can find a block by $n$ trails follows a geometric distribution:
\begin{align}\label{eqn:ntri1}
Pr(n {\rm{~trails~ used~ to~ mine~ a~ block}})=(1-p)^{n-1}p,
\end{align}
From \eqref{eqn:ntri1}, we have
\begin{align}\label{eqn:ntri2}
\nonumber Pr(n {\rm{~trails~ or ~fewer~ used~ to~ mine~ a~ block}})\\=\sum^{n-1}_{i=0}(1-p)^{n-1}p=1-(1-p)^n\xrightarrow{p\approx0} 1-e^{-np},
\end{align}
where  we use $e^{-1}=\lim\limits_{x\rightarrow\infty}(1-\frac{1}{x})^x$ for the approximation in the last equation when $D$ is very large.

Consider that a miner has the computation rate (i.e., hash rate) of $w$, i.e., the number of hash trails per unit time slot. If each trial for the miner takes the time of $\Delta t=1/w$, then $n$ trials take $t=n \Delta t$ amount of time. Given \eqref{eqn:ntri2} and $p=1/2^D$, we have
\begin{align}\label{eqn:ntri3}
\nonumber Pr(T\leq t)&=Pr(t {\rm{~amount~ of ~time~ or~ less~ to ~mine~ a~ block}})\\&=1-e^{-\frac{t}{\Delta t}\frac{1}{2^D}}=1-e^{-\frac{w}{2^D}t}=1-e^{-\lambda_Dt},
\end{align}
where $\lambda_D=w/2^D$.

From \eqref{eqn:ntri3}, the PDF of block generation time is
\begin{align}\label{eqn:ntri4}
p_T(t)=\frac{{\rm d}Pr(T\leq t)}{{\rm d}t}=\lambda_D e^{-\lambda_Dt}.
\end{align}
 According to \eqref{eqn:ntri4}, the block generation time follows an exponential distribution with rate of $\lambda_D=w/2^D$.

\section{PoA Protocol Design}\label{sec:prot}
In this section, we propose a modified PoW consensus protocol,  called \emph{Proof-of-Age} (PoA). Unlike in the native PoW protocol,  the miner in PoA benefits from its effective mining period (measured by \emph{Age-of-Work})  as long as it has devoted computing powers to maintaining the security of the blockchain, even if the miner has not successfully mined a block. %In this paper, we call the effective mining period as. %The PoA protocol  is detailed in the following.

\subsection{Hash Puzzle of PoA} \label{subsec:poa}
%We consider a blockchain network consisting of $N$ miners each with identical mining power $p$. The mining power $p$ can be treated as a unit of mining power. The total mining power of the network is $P=Np$.The miners can either mine the block independently and individually (i.e., solo mining) or pool their power (i.e., pooled mining).
%  %Mining pools can be formed by   a group of miners.
%For example, a mining pool can include $N_1$ miners and its mining power is then  $N_1p$. The design of PoA  ensures that there is no incentive to form the mining pools, i.e., the average profit per miner in the mining pool is less than the average profit of a solo miner.

\begin{figure*}[!h]
	\centering
	\includegraphics[height=1.8in]{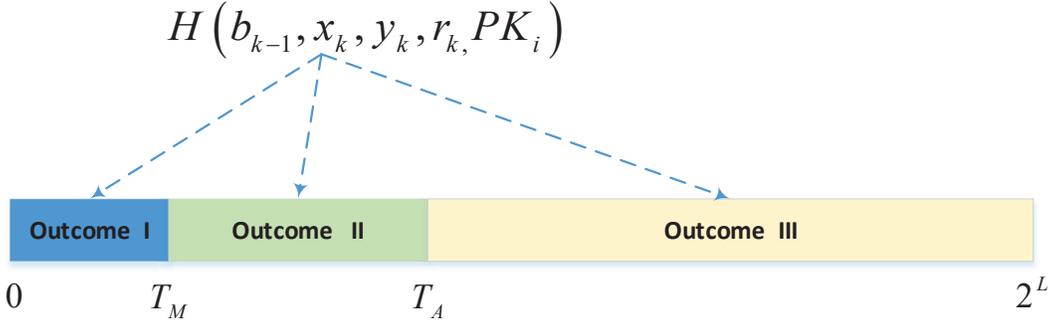}
	\caption{An illustration for three different mining outcomes.}
	\label{fig:result}
\end{figure*}

The miners can either mine the block independently and individually (i.e., solo mining) or mine the block together by aggregating their power into a pool (i.e., pooled mining). The design of PoA ensures that there is no incentive to form the mining pools, i.e., the average profit per miner in the mining pool is less than the average profit of a solo miner.

As in the PoW protocol, a new block in PoA is generated by solving a computational hash puzzle. The mining process of the original PoW protocol aims to solve the hash puzzle in \eqref{eqn:hash}. To introduce the metric of AoW to the mining process of PoA, we add a new parameter called \emph{age ring} into the input of the hash puzzle. For the exposition of our PoA protocol, we consider the mining process of a typical miner (either a mining pool or a solo miner) indexed by $i$ at the blockchain height $k$. At height $k$, miner $i$ executes the mining process by querying the hash of the following form:
\begin{align}\label{eqn:phash}
 H(b_{k-1}, x_k, y_k, r_k,PK_i),
\end{align}
where  $r_k$ is the age ring of the miner at blockchain height $k$, and $PK_i$ is the address of the miner (i.e., the hash of its public key). In \eqref{eqn:phash}, all the inputs are concatenated into a single string that is processed by the hash computation.

%already defined as  is the hash of  $y_k$ is defined in \eqref{eqn:hash}, $x_k$ denotes the solution string of the nonce,
 %, and all the inputs need to be included into the block for the mining of the block to be successful.

\subsection{Data Structure of Age Ring}\label{sub:data}
Suppose that miner $i$'s most recently mined block is at blockchain height $k-n$, where  $n>0$. The miner is allowed to compute a single age ring   at each blockchain height, and the age ring must be involved in the mining process, as expressed in \eqref{eqn:phash}. From its latest successful mining at blockchain height $k-n$  to the mining at blockchain height  $k$, miner $i$  has $n$  age rings  $\{r_k, r_{k-1}, \ldots, r_{k-n+1}\}$, where $r_l$  is the age ring of this miner attached to the block at blockchain height, $l=k-n+1, k-n+2, \ldots, k-1, k$. The data structure of the age ring $r_l$  attached to the block $l$   is given by
\begin{align}\label{eqn:age}
r_l =\left\{
          \begin{array}{lr}
 \langle r_{l-1}, b_{l-1}, w_l\rangle, ~~ l=k-n+2, \ldots, k-1, k  \\
\langle b_{l-1}, w_l\rangle,    ~~ ~~~~~~ l=k-n+1\\
             \end{array}
\right.
\end{align}
where $\langle\cdot\rangle$ is a concatenation operator and $b_{l-1}$ is the hash of previous block header at height $l-1$. Notably,  $w_l$ is the proof that the age ring $r_l$  is \emph{effective}, and this age-ring proof $w_l$  also needs to be  mined at height $l$. We will explain how to generate a proof for an {effective} age ring in Section \ref{sec:result}. From the data structure of age rings defined in \eqref{eqn:age}, we  find that the age ring $r_l$  at  height $l$  is chained to the  age ring $r_{l-1}$  at  height  $l-1$,  $l=k-n+2, \ldots, k-1, k$; and thus these age rings also form a chain which is started at height $l=k-n+1$  immediately after its latest successful mining at  height  $k-n$. Note that the  proof $w_l$  of age ring $r_l$  is initially set to as $w_l={\rm NULL}$\footnote{NULL is neither zero nor any other value, and it is an empty string that is of zero length and contains no value at all.}   at the very beginning of the mining process at  height $l$,  and it will be updated as explained later. Age rings are used to determine the AoW of the miner at each blockchain height.

\subsection{Computation of AoW}\label{sub:ug}

We denote the AoW of miner $i$  at  blockchain height $l$  by  $\Delta_i(l)$. As soon as the miner finds a block, its AoW immediately returns to the initial value of zero\footnote{If a miner is a newcomer that has never succeeded in the mining even once, its AoW is also set to be zero.}. Following the discussion in Section \ref{sub:data},  we have  $\Delta_i(k-n)=0$, due to the successful block generation of miner $i$ at blockchain height  $k-n$. Then, as the blockchain height increases from $k-n+1$  to  $l$, the AoW of miner $i$ at each height  $l$, i.e.,  $\Delta_i(l)$, is updated sequentially using the proof of age ring $r_l$  as
\begin{align}\label{eqn:aow}
\Delta_i(l)=\Delta_i(l-1)+I_{\neq {\rm NULL}}(r_l),
\end{align}
where  $l=k-n+1, \ldots, k-1, k$, and $I_{\neq {\rm NULL}}(r_l)$ is an indicator function defined by
\begin{align}\label{eqn:ind}
I_{\neq {\rm NULL}}(r_l) =\left\{
          \begin{array}{lr}
1,  ~~ r_l {\rm ~is ~ effective}  \\
0,    ~~ r_l {\rm ~is ~ ineffective}\\
             \end{array}
\right.
\end{align}
The AoW is used to adjust the mining difficulty in  the PoA protocol. Before delving into the adjustment of mining difficulty with AoW, we discuss how to obtain effective age rings and how to generate a block under PoA respectively.

\begin{figure*}[!h]
	\centering
	\includegraphics[height=1.8in]{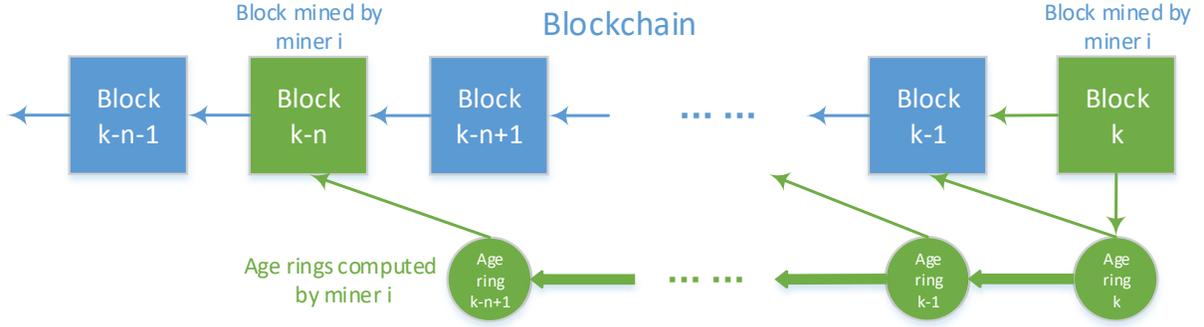}
	\caption{An illustration for the mining process under the PoA protocol.}
	\label{fig:block}
\end{figure*}

\subsection{Mining Outcomes of PoA}\label{sec:result}
Unlike in the PoW protocol, we set two mining targets $T_A$  and $T_M$  in the PoA protocol, where  $T_A$  and $T_M$  correspond to the mining of age ring and block respectively. Furthermore, we have  $T_M<T_A\leq 2^L$, where $L$ is the length of the outputs computed by the hash function  $H(\cdot)$. According to the ranges of the computed hash values, there are three different mining outcomes, as illustrated by the three bars with different colours in Fig. \ref{fig:result}. We present the details of the three outcomes under the PoA protocol, respectively.

\begin{itemize}
   \item \textbf{Outcome I}: the computed hash result is less than $T_M$, i.e.,
     \begin{align}\label{eqn:s1}
 H(b_{k-1}, x_k,  y_k,  r_k,PK_i)<T_M.
\end{align}
If this outcome happens, the miner obtains a valid nonce  $x_k$ as the solution for the new block at blockchain height  $k$. Note that the age ring $r_k$  also needs to be included into this new block together with the nonce, the hash of the previous block, and the transactions. Then the miner adds this block onto the blockchain and its AoW returns to zero, i.e., $\Delta_i(k)=0$, at this blockchain height. Therefore, the implication of mining target $T_M$  is the same as that of the mining target in PoW, i.e., it determines whether the miner finds a valid block. Moreover, we will adjust the mining target $T_M$  for a miner according to its AoW at each blockchain height, as explained in Section \ref{subsec:dif}.
   \item  \textbf{Outcome II}: the computed hash result is larger than $T_M$ but less than $T_A$, i.e.,
     \begin{align}\label{eqn:s2}
 H(b_{k-1}, x_k, y_k, r_k,PK_i)<T_A.
\end{align}
If this outcome happens, the corresponding nonce $x_k$  is not a solution for the new block, but this $x_k$ is a proof that the miner has devoted its computing power to the mining process at blockchain height $k$. Therefore, for this outcome, we say that the nonce $x_k$  becomes the proof that the age ring  $r_k$ is effective at this blockchain height, and the age-ring proof $w_k$   contained in the age ring  $r_k$ is updated to  $w_k=x_k$. Then, the AoW of this miner is increased by one, i.e., $\Delta_i(l)=\Delta_i(l-1)+1$, due to a factor that the age ring $r_k$  is effective now. After that, the miner includes the age ring $r_k$  into the input of the hash function to mine the   block. The mining target $T_A$  of age ring is fixed for all miners over all blockchain heights. The value of $T_A$  is preset according to the mining power unit  $p$. In this paper, we need to ensure that each miner with $p$ can compute an effective age ring that satisfies \eqref{eqn:s2} with probability approaching one in the mining process at each blockchain height.
   \item \textbf{Outcome III}: the computed hash result is larger than  $T_A$, i.e.,
\begin{align}\label{eqn:s3}
 H(b_{k-1}, x_k, y_k, r_k,PK_i)>T_A.
\end{align}
If this outcome happens, the miner fails to compute an effective age ring, not to mention the successful block mining. In this case, the age-ring proof keeps at its initial value  $w_k={\rm NULL}$, and the AoW of this miner at blockchain height $k$  is invariant.
 \end{itemize}

We illustrate the mining process of miner $i$ at blockchain height $k$ in Fig. \ref{fig:block}, where green squares represent the blocks mined by miner  $i$, blue squares represent the blocks mined by other miners, green rounds represent the age rings computed by miner  $i$, and arrows represent the dependency among blocks and age rings. This figure illustrates the case that mine $i$ has generated two blocks $k-n$  and  $k$  in sequence. Notably, the age rings of a miner also form a chain between any two adjacent blocks mined by this miner. Since the mining of block $k$ is performed based on the hash of the previous block header $b_{k-1}$ and age ring $r_k$  (as specified by the hash form in \eqref{eqn:phash}), there are two arrows coming out of block $k$ that point to block $k-1$ and age ring  $r_k$, respectively. Since age ring $r_l$ links to $r_{l-1}$ through the recursive data structure of age rings (as defined in \eqref{eqn:age}), there is an arrow coming out of age ring $r_l$ that points to age ring  $r_{l-1}$. Meanwhile, the computation of age ring $r_l$ also incorporates the hash of block header $b_{l-1}$,  and thus these is also an arrow coming out of $r_l$ pointing to the block at height $l-1$.  We stress that i) the AoW of miner $i$ returns to zero at the blockchain height where it has successfully generated the block; ii) AoW keeps increasing with the generation of effective age rings.

 \begin{figure*}[!h]
	\centering
	\includegraphics[height=1.8in]{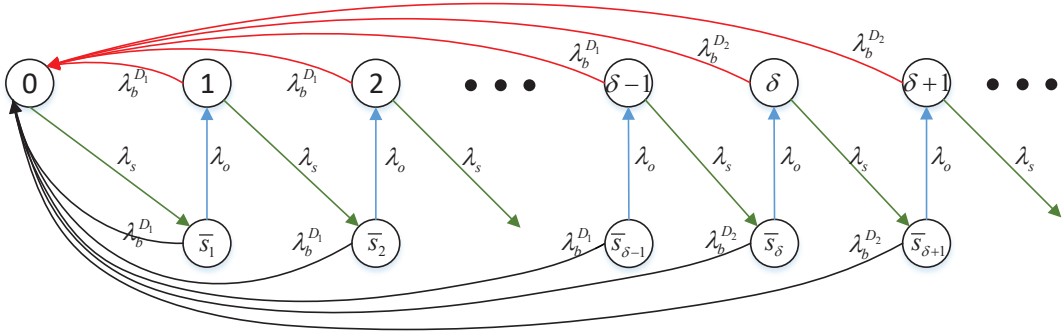}
	\caption{CTMC of the block arrival process under PoA.}
	\label{fig:ctmc}
\end{figure*}

\subsection{Difficulty Adjustment}\label{subsec:dif}
The essence of our PoA protocol is to construct a function  $f(\cdot)$ that maps the AoW of each miner, i.e., $\Delta_i(l)$, to a mining difficulty level  $D=f(\Delta_i(l))\in\{d_1, d_2, \ldots, d_w, \ldots\}$, where $d_m$  denotes the mining difficulty level when  $\Delta_i(l)=m$. In this paper, we consider a case that the difficulty level $D$  can only take two values, i.e.,  $D\in\{D_1, D_2\}$, where  $D_1\gg D_2$. Specifically, the function $f(\cdot)$ that maps the AoW mining difficulty levels is given by
          \begin{align}\label{eqn:adj}
D =f(\Delta_i(l)) =\left\{
          \begin{array}{lr}
D_1,  ~~ \Delta_i(l)<\delta  \\
D_2,    ~~ \Delta_i(l)\geq\delta\\
             \end{array}
\right.
\end{align}
This mapping of mining difficulty level means that if the AoW is no less than a threshold  $\delta$, the miner mines with a larger mining difficulty  $D_1$; otherwise, the miner mines with a smaller difficulty  $D_2$. From \eqref{eqn:hash}, we have  $T(D)=2^{L-D}$, which means that a larger $D$  gives a smaller  $T_M$. Recall that the mining target of age ring $T_A$ is preset and fixed for all miners. The general form of the multi-level mining difficulty mapping function $f(\cdot)$  will be considered in the future work. Regarding our PoA protocol, we have the following remarks.

\begin{remark}[Age-ring chain] The idea on the mining process in \eqref{eqn:phash} and the recursive data structure of age rings in \eqref{eqn:age} is that all age rings between two adjacent blocks generated by the same miner form a chain, i.e., the current age ring is computed using the previous age ring. Due to the chain of age rings, the age rings cannot be computed at the same time and can only be computed one by one along with each block. Thus, the miner has no incentive to split its power into multiple parts and use these parts to simultaneously generate multiple age rings in each mining round. %The miner has to mine the blocks one by one using its computing power.
%Notably, the age rings computed by a miner cannot be used by other miners, since the age rings are computed with the address of the miner.
\end{remark}

\begin{remark}[Negligible overhead of age ring] If the miner has generated a block (i.e., finding a nonce satisfying \eqref{eqn:age}), it needs to include the eligible nonce together with current age ring into the block. Note that the age ring is just a set of hashes; hence the storage overhead introduced by the age rings is negligible.
\end{remark}

\begin{remark}[Block verification] Once a miner successfully generates a block, it will include the nonce, the age ring, and its address into the block, and broadcast the block to the whole network. The network can easily validate this block by checking if the nonce satisfies \eqref{eqn:age} and the age rings satisfy \eqref{eqn:s2}, respectively.
\end{remark}

\begin{remark}[Sybil attack resistance] Each miner $i$ on the blockchain is identified by its address $PK_i$  and the age rings are tied to the address. It is impossible that a miner can mine the block using the age rings obtained by other miners with different addresses. If a newcomer (i.e., a miner with a new address) begins to mine, its AoW is defaulted to be zero. This mechanism can prevent malicious miners from making multiple addresses to carry out sybil attacks, since each new address needs to mine with the largest difficulty.
\end{remark}

\begin{remark}[Memoryless mining]\label{rem:mem} When a miner generates an effective age ring at a blockchain height before any other miner successfully generates a block, this miner can include this effective age ring into \eqref{eqn:phash} and continue to mine the block at this height. Due to the memoryless property of the mining process \cite{rosenfeld2011analysis}, the block generation probability of this miner after an effective age ring is found is still the same as the probability when it mines at the very beginning at this  height.
\end{remark}

%To investigate the average throughput of PoA, we formulate the mining process under PoA as a continuous-time Markov chain in the following section.
\section{Model Analysis}\label{sec:model}
In this section, we mathematically model the mining process of PoA and analyze the block generation rate (i.e., block throughput) of the PoA based blockchain.

\subsection{CTMC Model}\label{subsec:ctmc}

We consider a blockchain network consisting of $N$ miners each with unit computing power. The total mining power of the network is $N$. Suppose that in the network, there exists a single mining pool that aggregates $N_1$ miners, and each of the other $N_2=N-N_1$ miners performs the solo mining. The power of the pool is $N_1$, i.e., the sum power of its members.

The investigation on the block throughput of the PoA blockchain is related to the stochastic process of block arrival on the blockchain. The average rate of block arrival process is impacted by the hash powers and the  mining difficulties of the miners. According to our PoA protocol, the mining difficulty of each miner is determined by its AoW that changes with the interactions among different miners in the mining process. Consequently, the AoW states of the miners are correlated and coupled together. In this context, we need to track the AoW states of the miners for the block throughput analysis.

The most accurate way to analyze the block throughput of the PoA blockchain is to model the AoW states of each individual miner. However, this will lead to unaffordable model complexity due to the huge AoW state space for handling all of the miners. Specifically, the number of AoW states of all the miners is exponentially related to the total number of the miners in the network. To address this issue, we propose to model the mining process of PoA with a {\em decoupling} hypothesis, where we focus on the mining process of a typical miner and treat the aggregation of all other miners as the surroundings of this typical miner. Let us define one other miner with respect to the typical miner as an exterior miner. Under this decoupling hypothesis, we only need to model the interior states of the typical miner and the interaction between the typical miner and its surroundings, i.e., the aggregation of all  exterior miners.

The interior state of the typical miner will transit as long as either it finds a block/an effective age ring or any exterior miner finds a block. Given fixed mining difficulty level and AoW, the inter-arrival time of blocks (i.e., the block generation time) found by a miner follows an exponential distribution. Moreover, since the generation of effective age rings is also made from the mining process, the inter-arrival time of effective age rings found by the typical miner also follows an exponential distribution. However, since the AoW states of different exterior miners are diverse and correlated, different exterior miners may have different mining difficulties, and the mining processes of different exterior miners follow exponential distributions with different rates. As a result, it is challenging to characterize the block inter-arrival time of all  exterior miners. Fortunately, as we will show in Section \ref{subsec:dis}, the block inter-arrival time of all exterior miners can be well approximated by a single exponential distribution. Now, since the inter-arrival time of each event that triggers the state transitions under PoA can be treated as an exponential distribution, we can model the mining process of a typical miner under the PoA protocol using the tool of Continuous-Time Markov Chain (CTMC) \cite{anderson2012continuous}, described as follows.

Fig. \eqref{fig:ctmc} shows the proposed CTMC model under the PoA protocol consisting of two types of states:
\begin{itemize}
  \item State $i$,  $i\in\{0,1,\ldots, \delta, \delta+1,\ldots\}$, means that the current AoW value of this typical miner is  $i$. The effective age ring has  not been generated at the current blockchain height, and thus it can be generated;
  \item State $\bar{s}_i$,  $\bar{s}_i\in \{\bar{s}_1, \bar{s}_2,\ldots, \bar{s}_\delta, \bar{s}_{\delta+1}, \ldots\}$, means that the current AoW value of this typical miner is  $i$. The effective age ring has been generated at the current blockchain height, and thus it is not allowed to be generated again.
\end{itemize}

Assume that the typical miner of interest is in state  $i$. Following the PoA protocol, the typical miner attempts to mine the block by conducting the hash queries in \eqref{eqn:phash}. The mining result that will trigger the state transition of the miner is either a nonce satisfying \eqref{eqn:s1} or a nonce satisfying \eqref{eqn:s2}. If a nonce that satisfies \eqref{eqn:s1} is found, a new block is found by this typical miner and then the state of this miner transits to state  $0$. If a nonce that satisfies \eqref{eqn:s2} is found, an effective age ring is generated by the typical miner and the state of this miner transits to state  $\bar{s}_i$, where the miner updates its AoW using the effective age ring and adjusts its mining difficulty with  updated AoW accordingly. Note that after the miner generates the effective age ring at the current blockchain height, it continues to mine the block at this height. As a result, the typical miner can either generate a block ahead of all exterior miners in the current mining round, or increase its AoW for the successor mining rounds if any exterior miner generates the block ahead. Since the AoW value of the typical miner drops to zero (i.e., the state of this miner transits to zero) if the block is generated by the typical miner, the recurrence time of state 0 is the block inter-arrival time of the typical miner (i.e., the time between two successive visits of state 0). In view of these transitions, we define the transition rates in the CTMC as{\footnote{It is customary not to show the self-transition from state $i$ to state $i$ in the CTMC, since the transition rate can be cancelled out in the balance equation at equilibrium for state $i$.}}
\begin{itemize}
  \item effective age ring generation rate under difficulty level $D_s$, i.e., $\lambda_s$;
  \item block generation rate of the typical miner under difficulty level $D_1$, i.e., $\lambda^{D_1}_b$;
  \item block generation rate of the typical miner under difficulty level $D_2$, i.e., $\lambda^{D_2}_b$;
  \item block generation rate of all exterior miners, i.e., $\lambda_o$.
\end{itemize}
where $\lambda^{D_1}_b=w/2^{D_1}$, $\lambda^{D_2}_b=w/2^{D_2}$, and $\lambda_s=w/2^{D_s}$. Due to the memoryless property of exponential distributions, the transition rate from states $i$ to $\bar{s}_i$  is independent of that from states $i+1$ to $0$. From Remark \ref{rem:mem}, if the typical miner has generated an effective age ring, he does not become any closer to the block generation than he was in the beginning.

\subsection{Block Throughput Analysis}
Let $P_i$ denote the steady-state probability that the chain is in state $i$, $i\in\{0,1,\ldots, \delta, \delta+1,\ldots\}$. By equating the rates of flow into and out of each state, the steady-state probabilities in the irreducible{\footnote{A Markov chain is irreducible if every state can be reached from every other state.}}  CTMC yield
\begin{subequations}
\begin{align}
&(\lambda^{D_1}_b+\lambda_s)P_i=\lambda_oP_{\bar{s}_i}, 1\leq i\leq \delta-1,\\
&(\lambda^{D_2}_b+\lambda_s)P_i=\lambda_o P_{\bar{s}_i},  i\geq \delta, \\
&(\lambda^{D_1}_b+\lambda_o)P_{\bar{s}_i}=\lambda_s P_{i-1},  1\leq i\leq \delta-1, \\
&(\lambda^{D_2}_b+\lambda_o)P_{\bar{s}_i}=\lambda_s P_{i-1},  i\geq \delta, \\
\label{eqn:sum}
&\sum^{\infty}_{i=0}P_i+\sum^{\infty}_{i=1}P_{\bar{s}_i}=1.
\end{align}
\end{subequations}

By some mathematical manipulation, we have
        \begin{subequations} \begin{align}\label{eqn:c1}
P_i &=\left\{
          \begin{array}{lr}
P_0(\Phi_1)^i,  1\leq i\leq \delta-1  \\
P_0(\Phi_1)^{\delta-1}(\Phi_2)^{i-(\delta-1)}, i\geq \delta\\
             \end{array}
\right.\\ \label{eqn:c2}
P_{\bar{s}_i} &=\left\{
          \begin{array}{lr}
P_0\frac{\lambda_s}{(\lambda^{D_1}_b+\lambda_o)}(\Phi_1)^{i-1},  1\leq i\leq \delta-1  \\
P_0\frac{\lambda_s}{(\lambda^{D_2}_b+\lambda_o)}(\Phi_1)^{\delta-1}(\Phi_2)^{i-1-(\delta-1)}, i\geq \delta\\
             \end{array}
\right.
\end{align}  \end{subequations}
where $\Phi_1=\frac{\lambda_s\lambda_o}{(\lambda^{D_1}_b+\lambda_s)(\lambda^{D_1}_b+\lambda_o)}$ and  $\Phi_2=\frac{\lambda_s\lambda_o}{(\lambda^{D_2}_b+\lambda_s)(\lambda^{D_2}_b+\lambda_o)}$.

Plugging \eqref{eqn:c1} and \eqref{eqn:c2} into \eqref{eqn:sum}, we have
 \begin{align}\label{eqn:c3}
\nonumber &P_0=1-\sum^{\infty}_{i=1}(P_i+P_{\bar{s}_i})\\
&=\frac{1}{1+\frac{\lambda_s}{\lambda^{D_1}_b}+(\frac{\lambda_s\lambda_o}{(\lambda^{D_1}_b+\lambda_s)(\lambda^{D_1}_b+\lambda_o)})^{\delta-1}(\frac{\lambda_s}{\lambda^{D_2}_b}-\frac{\lambda_s}{\lambda^{D_1}_b})}.
\end{align}

Given \eqref{eqn:c1}, \eqref{eqn:c2}, and \eqref{eqn:c3}, the block generation rate of the typical miner is given by
 \begin{align}\label{eqn:typ}
\nonumber &\rho_{\rm typical}=\sum^{\delta-1}_{i=0}(P_i+P_{\bar{s}_i})+\sum^{\infty}_{i=\delta}(P_i+P_{\bar{s}_i})=(\lambda^{D_1}_b+\lambda_s)P_0\\
&=\frac{1}{\frac{1}{\lambda^{D_1}_b}+(\frac{\lambda_s}{\lambda^{D_1}_b+\lambda_s})^{\delta}(\frac{\lambda_o}{\lambda^{D_1}_b+\lambda_o})^{\delta-1}(\frac{1}{\lambda^{D_2}_b}-\frac{1}{\lambda^{D_1}_b})},
\end{align}
where $\rho_{\rm typical}=\lambda^{D_1}_b$ if either $\delta$ goes to infinity (i.e., no one can reach state $\delta$) or $D_1=D_2$ (i.e., the difficulty control of PoW), and $\rho_{\rm typical}=\lambda^{D_2}_b$ if $\delta=1$ (i.e., the difficulty reduces as long as a single effective age ring is generated). From \eqref{eqn:typ}, we further have $\lambda^{D_1}_b\leq\rho_{\rm typical}\leq \lambda^{D_2}_b$, since $\lambda^{D_1}_b\leq\lambda^{D_2}_b$.

For a miner in the PoW based blockchain network, the block generation rate of the miner is the ratio of its computing power to the total computing power of the whole network; moreover, the variance of block inter-arrival time is proportional to the inverse of its computing power \cite{rosenfeld2011analysis}. Therefore, the block inter-arrival time usually exhibits very large variance for a solo miner  with relatively small computing power. As a result, it may take extremely long time for the solo miner to successfully mine a block. In reality, this result incentivizes  the solo miners to aggregate their computation powers into a mining pool, where all participating miners work together to mine and share the revenue as long as any miner in the pool successfully finds a block. With this approach of pooled mining, the miners in the pool can greatly reduce their variances of the block inter-arrival time and get paid regularly in every day, although the block generation rate of each miner in the pool is not changed. The pooled mining can indeed bring significant benefits to the miners; however, the pooling of computing power undermines the decentralized nature of the blockchain, which can potentially cause serious security threats to the blockchain.

In the following, we will reveal that the PoA protocol   discourages the pooled mining, and thus it is not worthwhile for the miners to form the mining pool. To this end, we first characterize the block generation rates of the mining pool and all exterior miners (i.e., solo miners) based on \eqref{eqn:typ}, and then prove that the block generation rate of the mining pool under PoA is not proportional to its aggregated computing power and can even be much less than its proportion.

Let us use the CTMC in Fig. \ref{fig:ctmc} to track the mining process of the mining pool.
 %if $N_1$ is not too large and $N_1$ is far less than $N$.
In this case, the typical miner becomes the mining pool, and the decoupling assumption still holds. Using \eqref{eqn:typ}, we can specify the block generation rates of the mining pool and each solo miner respectively as
  \begin{align}\label{eqn:pool}
  &\rho_{\rm pool}=\frac{N_1}{2^{D_1}+\phi_{\rm pool}},\\
 \label{eqn:solo}
  &\rho_{\rm solo}=\frac{1}{2^{D_1}+\phi_{\rm solo}},
\end{align}
 where
 \begin{align}\label{eqn:pool1}
  \nonumber \phi_{\rm pool}=&(\frac{2^{D_1+D_2}}{2^{D_2+D_s}+2^{D_1+D_2}})^\delta
(\frac{N_2\rho_{\rm solo}}{N_1/2^{D_1}+N_2\rho_{\rm solo}})^{\delta-1}\times \\&(2^{D_2}-2^{D_1}),\\
 \label{eqn:solo1}
\nonumber \phi_{\rm solo} =&(\frac{2^{D_1+D_2}}{2^{D_2+D_s}+2^{D_1+D_2}})^\delta\times\\&
(\frac{(N_2-1)\rho_{\rm solo}+\rho_{\rm pool}}{1/2^{D_1}+(N_2-1)\rho_{\rm solo}+\rho_{\rm pool}})^{\delta-1}(2^{D_2}-2^{D_1}),
\end{align}
where, for the pool in \eqref{eqn:pool}, the aggregated block generation rate of all exterior miners is  $\lambda_o=N_2\rho_{\rm solo}$; for each solo miner in \eqref{eqn:solo}, the aggregated block generation rate of all exterior miners is  $\lambda_o=(N_2-1)\rho_{\rm solo}+\rho_{\rm pool}$. Using the fixed-point iteration (FPI)  in Algorithm 1, we can compute the block generation rates of the mining pool and each solo miner.

From \eqref{eqn:pool} and \eqref{eqn:solo},   total throughput of the PoA based blockchain network is given by
 \begin{align}\label{eqn:total}
 \rho_{\rm total}=\rho_{\rm pool}+N_2\rho_{\rm solo}.
\end{align}

Furthermore, the relative rate gain of the pool over all exterior miners is defined as the ratio between the block generation rates of the pool and all solo miners, given by
 \begin{align}\label{eqn:gain}
 F_{\rm PoA}=\frac{\rho_{\rm pool}}{N_2\rho_{\rm solo}}=\frac{N_1}{N_2}G,
\end{align}
where
 \begin{align}\label{eqn:g}
 G=\frac{{2^{D_1}+\phi_{\rm solo}}}{{2^{D_1}+\phi_{\rm pool}}}.
\end{align}
Furthermore, it is clear that $F_{\rm PoW}=N_1/N_2$  is the relative rate gain under PoW. As such, we refer to $G$ as the gap of relative rate gains between PoA and PoW. From \eqref{eqn:g}, it can be verified that $G=1$ holds if either $N_1=1$ (i.e., no pooled mining) or $\delta$ goes to infinity. In this case,  $F_{\rm PoW}=F_{\rm PoA}$.

To show that PoA can discourage the pooled mining, we first verify that $G< 1$, i.e.,
 \begin{align}\label{eqn:gs1}
F_{\rm PoA}=\frac{\rho_{\rm pool}}{N_2\rho_{\rm solo}}< F_{\rm PoW}=\frac{N_1}{N_2}.
\end{align}
To prove \eqref{eqn:gs1}, it is equivalent to proving that $\frac{(N_2-1)\rho_{\rm solo}+\rho_{\rm pool}}{1/2^{D_1}+(N_2-1)\rho_{\rm solo}+\rho_{\rm pool}}> \frac{N_2\rho_{\rm solo}}{N_1/2^{D_1}+N_2\rho_{\rm solo}}$. The proof is straightforward and omitted here.

\begin{figure*}[!h]
	\centering
	\includegraphics[height=2.8in]{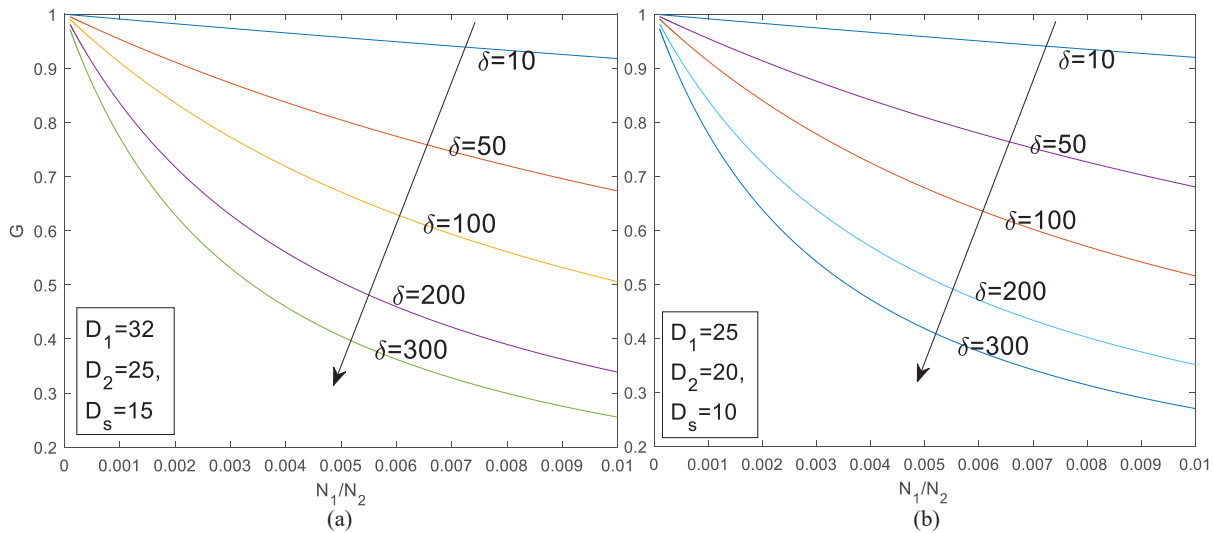}
	\caption{The gap of relative throughput gain between PoA and PoW: (a) when $D_1=32$, $D_2=25$, $D_3=15$; (a) when $D_1=25$, $D_2=20$, $D_3=10$.}
	\label{fig:gap}
\end{figure*}

Second, the numerical results in Fig. \ref{fig:gap} show that the gap $G$ in \eqref{eqn:g} can be very large when $\delta$ goes larger within a finite range. For example, it is observed in Fig. \ref{fig:gap}(a) that the gap can be reduced to 25$\%$ when $\delta=300$ under $D_1=32, D_2=25, D_s=15$, and $\frac{N_1}{N_2}\in [10^{-4}, 10^{-2}]$.

Therefore, we conclude that the pooling under PoA can disincentivize the pooled mining from the block generation rate perspective.

\begin{algorithm}[t]
	\setstretch{0.9}
	\caption{Fixed-point iteration}\label{alg:1}\begin{small}
		\begin{algorithmic}[1]
			\State {Initialization: $N, N_1, N_2, \delta, D_1, D_2, D_s, \epsilon, \{\rho_{\rm pool}(1), \rho_{\rm solo}(1)\}=\{N_1/2^{D_1}, 1/2^{D_1}\}$}
			\While{$\gamma>\epsilon$}
			\State {$k\leftarrow k+1$}
			\State {Update $ \{\rho_{\rm pool}(k), \rho_{\rm solo}(k)\}$ by plugging $ \{\rho_{\rm pool}(k-1), \rho_{\rm solo}(k-1)\}$ into \eqref{eqn:pool} and \eqref{eqn:solo}}
			\State {$\gamma=\frac{(\rho_{\rm pool}(k)-\rho_{\rm pool}(k-1))^2+(\rho_{\rm solo}(k)-\rho_{\rm solo}(k-1))^2}{(\rho_{\rm pool}(k-1))^2+(\rho_{\rm solo}(k-1))^2}$}
			\EndWhile
			\State {Output: $\{\rho_{\rm pool}(k), \rho_{\rm solo}(k)\}$}
	\end{algorithmic}\end{small}
\end{algorithm}

\subsection{Distributions of Block Inter-Arrival Time}\label{subsec:dis}

%In the above, we used the result that the block inter-arrival time of the mining pool is exponentially distributed and the block inter-arrival time of the solo miners is Gaussian distributed to analyze the block generation rates.

In this part, we analytically validate the distributions of block inter-arrival times of the mining pool and the solo miner respectively. The key point of the following analysis is the relative magnitudes of the ``time it takes for the state 0 to reach $\delta$'' and the ``time to generate a block at difficulty level of $D_1$''. If the former is much shorter than the latter, then the effective difficulty level is $D_2$.

First, let us consider a solo miner. For the solo miner, the average time to generate an effective age ring is  $2^{D_s}$, and the average time to move from state $\bar{s}_i$  to state $i$ is less than  $2^{D_1}/(N-1)$. In total, the average time from state $0$ to state $\bar{s}_{\delta}$  is therefore less than  $\delta(2^{D_s}+\frac{2^{D_1}}{N-1})$. Note that the average time to generate a block at difficulty level $D_1$ is $2^{D_1}$. In our setting, $N$ is large and $D_1\gg D_s$. Thus, the state almost always reaches $\bar{s}_\delta$  before a block is generated at $D_1$. Thus, as an approximation, we can treat the effective difficulty level of a solo miner as $D_2$. Its average block generation time is therefore exponentially distributed with rate of  $1/2^{D_2}$.

Next, consider the mining pool. The average time to generate an effective age ring (go from state $i$ to state  $\bar{s}_{i+1}$) is  $2^{D_s}/N_1$. With the approximation in the previous paragraph, the average time to go from state $\bar{s}_{i}$  to state $i$ is  $1/\lambda^{D_2}_b=2^{D_2}/N_2$. In total, the average time to go from state $0$ to state $\bar{s}_{\delta}$  is therefore  $\bar{t}_\delta=\delta(2^{D_s}/N_1+2^{D_2}/N_2)$. The associated random variable is a sum of  $2\delta$ random variables. Using the law of large numbers, we can approximate the generation time of $\bar{s}_{\delta}$  as a Gaussian distribution with mean $\bar{t}_\delta$  and variance $\sigma^2_\delta=\bar{t}_\delta/\delta$ (i.e., $t_\delta\sim \mathcal{N}(\bar{t}_\delta, \sigma^2_\delta)$), when $\delta$  is sufficiently large. In addition, the average time for the mining pool to generate a block at difficulty level of $D_1$ is  $1/\lambda^{D_1}_b=2^{D_1}/N_1$.

Let  $t_\delta$ be a realization of the time to reach state $\bar{s}_{\delta}$. Let $t$ be the block inter-arrival time. Given  $t_\delta$, we have the conditional PDF of the block inter-arrival time of the pool as
    \begin{align}\label{eqn:pdf}
p(t|t_\delta)  =\left\{
          \begin{array}{lr}
\lambda^{D_1}_be^{-t\lambda^{D_1}_b}u(t_\delta-t), ~~~~~~~~~~~~~~~  {\rm for~} t<t_\delta  \\
\lambda^{D_2}_be^{-(t-t_\delta)\lambda^{D_2}_b}e^{-\lambda^{D_1}_bt_\delta}u(t-t_\delta),    {\rm for~} t\geq t_\delta \\
             \end{array}
\right.
\end{align}
where the second equation implies that the pool generates the block at difficulty level $D_2$ if $t\geq t_\delta$, and $e^{-\lambda^{D_1}_bt_\delta}$ denotes the probability that the pool cannot find a block within time  $t_\delta$. Recall that  $t_\delta\sim \mathcal{N}(\bar{t}_\delta, \sigma^2_\delta)$. Taking average of  $t_\delta$, the PDF of the block inter-arrival time is given by
    \begin{align}\label{eqn:apdf}
p(t) =p_1(t)+p_2(t)=\int_{\delta} p(t|t_\delta)p(t_{\delta}){\rm d}t_{\delta}, \forall t\geq 0,
\end{align}
where
\begin{subequations}
   \begin{align}\label{eqn:apdf1}
 \nonumber p_1(t)&=\lambda^{D_1}_be^{-t\lambda^{D_1}_b}
 \int^{\infty}_{t}\frac{e^{-\frac{(t_\delta-\bar{t}_\delta)^2}{2\sigma^2_\delta}}}{\sqrt{2\pi\sigma^2_\delta}}{\rm d}t_{\delta}\\
 &=\lambda^{D_1}_be^{-t\lambda^{D_1}_b}Q\left(\frac{t-\bar{t}_\delta}{\sigma_\delta}\right), {\rm ~for~} t<t_\delta,\\
 \nonumber p_2(t)&=\lambda^{D_2}_b
 \int^{t}_{-\infty}e^{-(t-t_\delta)\lambda^{D_2}_b}e^{-\lambda^{D_1}_bt_\delta}\frac{e^{-\frac{(t_\delta-\bar{t}_\delta)^2}{2\sigma^2_\delta}}}{\sqrt{2\pi\sigma^2_\delta}}{\rm d}t_{\delta} \\
 \nonumber &=\lambda^{D_2}_be^{-t\lambda^{D_2}_b}e^{\frac{\eta(2\bar{t}_\delta+\lambda^{D_1}_b\sigma^2_\delta)}{2}}\left(1-Q(\frac{t-(\bar{t}_\delta+\eta\sigma^2_\delta)}{\sigma_\delta})\right),\\
  &~~~{\rm ~for~} t\geq t_\delta,
\end{align}
\end{subequations}
with $Q(\cdot)$ being the $Q$-function and $\eta=\lambda^{D_2}_b-\lambda^{D_1}_b$. Fig. \ref{fig:simdis}  in Section \ref{sec:sim} will demonstrate the distributions of the mining pool and solo miners.

\begin{figure*}[!ht]
	\centering
	\includegraphics[height=3in]{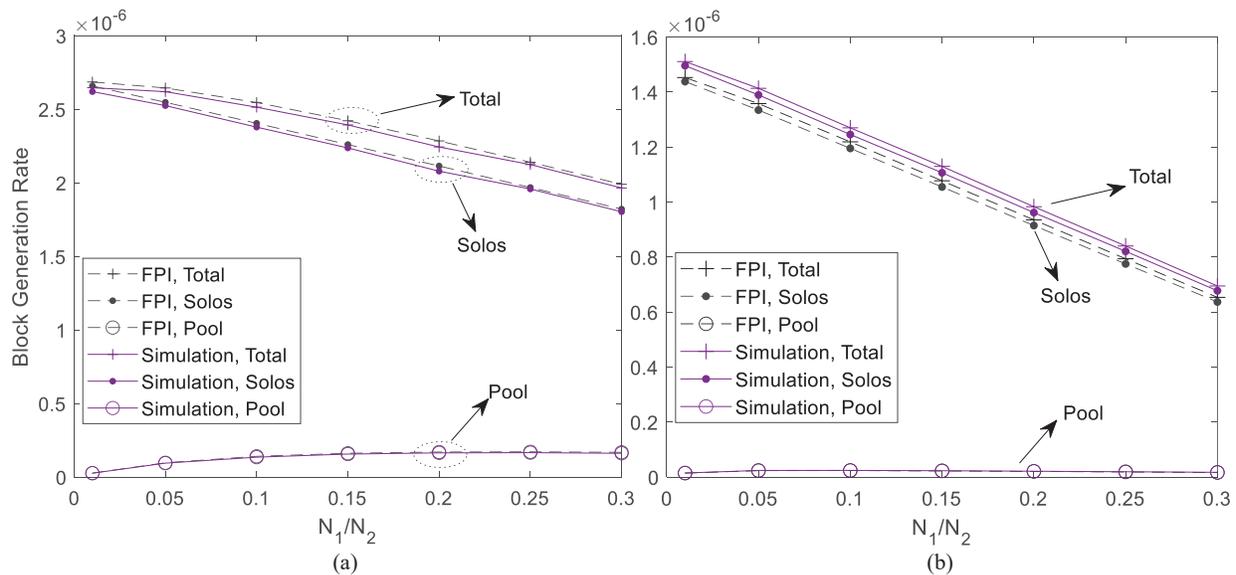}
	\caption{Throughputs under FPI and simulations with (a) $\delta=10$ and (b)  $\delta=50$.}
	\label{fig:simFPI}
\end{figure*}

\section{Numerical and Simulation Results }\label{sec:sim}
We use event-driven simulations to model the block generation process under PoA, where an event represents a state of the CTMC in Fig. \ref{fig:ctmc}. With the event-driven simulations, we only focus on the time that a state holds before it moves to the other state. In this way, we can record the block inter-arrival times of both mining pool and  solo miner.

\subsection{Simulations of Throughput and Block Inter-arrival Time}

Consider that a blockchain network with $N=100$ miners for $10^6$ blocks, where $D_1=32, D_2=25$, and $D_s=15$.

Fig. \ref{fig:simFPI} shows  total throughput of the PoA based blockchain network  with $\delta=10$  and  $\delta=50$, respectively. Consider $N_1/N_2\in [1/100, 1/3]$. The numerical results of total throughput of the blockchain network and the block generation rates of mining  pool and solo miners are obtained from the FPI in Algorithm \ref{alg:1}. First, it is observed that the numerical results match the simulation results. Second, each subfigure shows that the throughput decreases as the pool becomes larger. Compared with Fig. \ref{fig:simFPI}(a), both total throughput and block generation rates in Fig.  \ref{fig:simFPI}(b) reduce due to a higher $\delta$.

\begin{figure*}[t]
  \centering
        \includegraphics[height=5in]{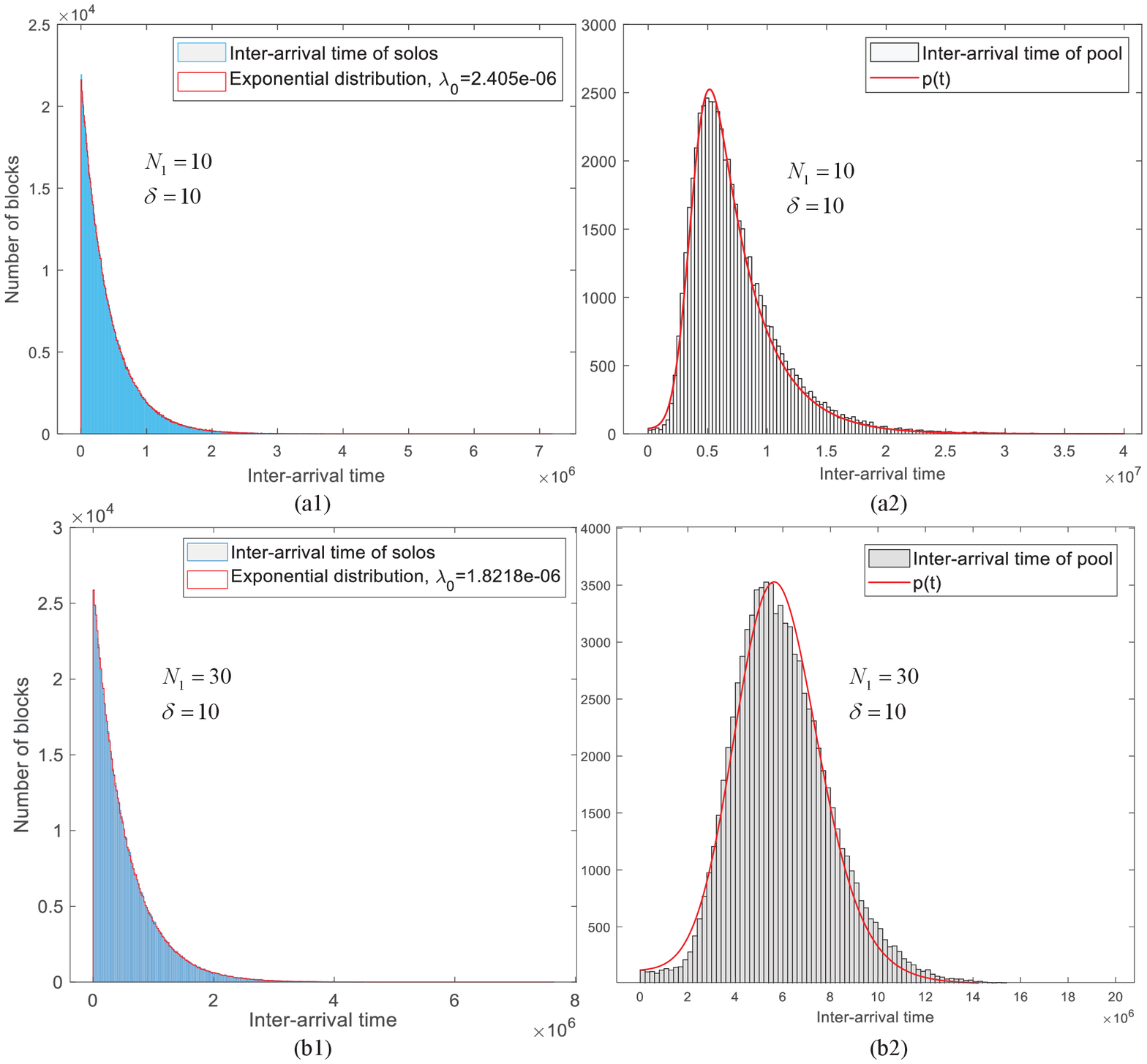}
       \caption{Block inter-arrival times of the pool and solo miners with (a) $N_1=10$ (b) $N_1=30$.}
        \label{fig:simdis}
\end{figure*}

Fig. \ref{fig:simdis} plots the histograms of block inter-arrival times of the mining pool and solo miners under  $\delta=10$. The height of each bar in the histogram plot indicates the number of block inter-arrival times that fall into the corresponding range. Figs. \ref{fig:simdis}(a) and \ref{fig:simdis}(b) consider that $N_1=10$ and $N_1=30$ respectively. For reference purpose, we also numerically plot the exponential distributed curves with rate of $\lambda_o=N_2\rho_{\rm solo}$  obtained by the FPI in Algorithm \ref{alg:1} and the derived  distribution of $p(t)$ in Section \ref{subsec:dis}. First, it is observed from Figs. \ref{fig:simdis}(a1) and \ref{fig:simdis}(b1) that the block inter-arrival times of all solo miners follow the exponential distributions with rates of  $\lambda_o=N_2\rho_{\rm solo}$, $N_2=90$ and $N_2=70$, respectively. This validates our assumption in the formulation of CTMC in Section \ref{subsec:ctmc}. Second, Figs. \ref{fig:simdis}(a2) and \ref{fig:simdis}(b2) show the distributions of block inter-arrival time of the mining pool follow our analytical result in Section \ref{subsec:dis}.

 \subsection{Comparison of block generation rates between PoW and PoA}
Consider that $D_1=32, D_2=25, D_s=10, N=10^6, \delta=100$. Let the number of miners in the pool be $N_1=rN$, where $r$ denotes the power ratio of the pool over the entire network. For each power ratio, the same total block throughput is fixed for both PoA and PoW based blockchain networks. The block generation rate of each individual miner is normalized according to the total block throughput,  and the vertical error bar is determined by the standard deviation.

\begin{figure}[!h]
  \centering
        \includegraphics[height=0.75\columnwidth]{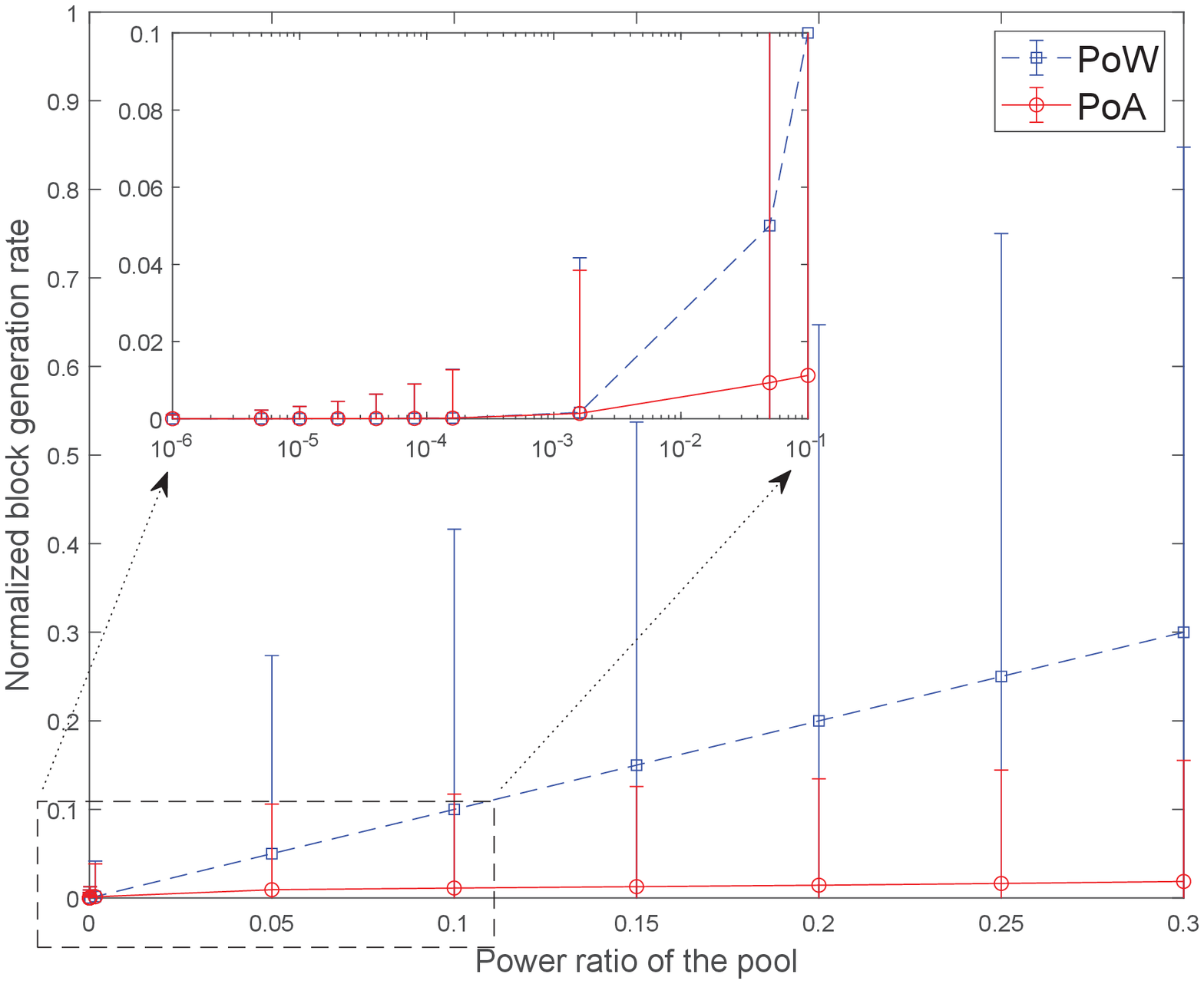}
       \caption{Comparison of normalized block generation rates of the pool under PoW and PoA.}
        \label{fig:simpool}
\end{figure}

Fig. \ref{fig:simpool} shows the normalized block generation rates of the pool under PoW and PoA respectively. First, as the power ratio goes up, the block generation rate of the pool under PoW linearly increases. Second, the increase of block generation rate of the pool under PoA is negligible. Third, the variance of block generation rate of the pool under PoA is much less significant than that of PoW.

\begin{figure}[!h]
  \centering
        \includegraphics[height=0.8\columnwidth]{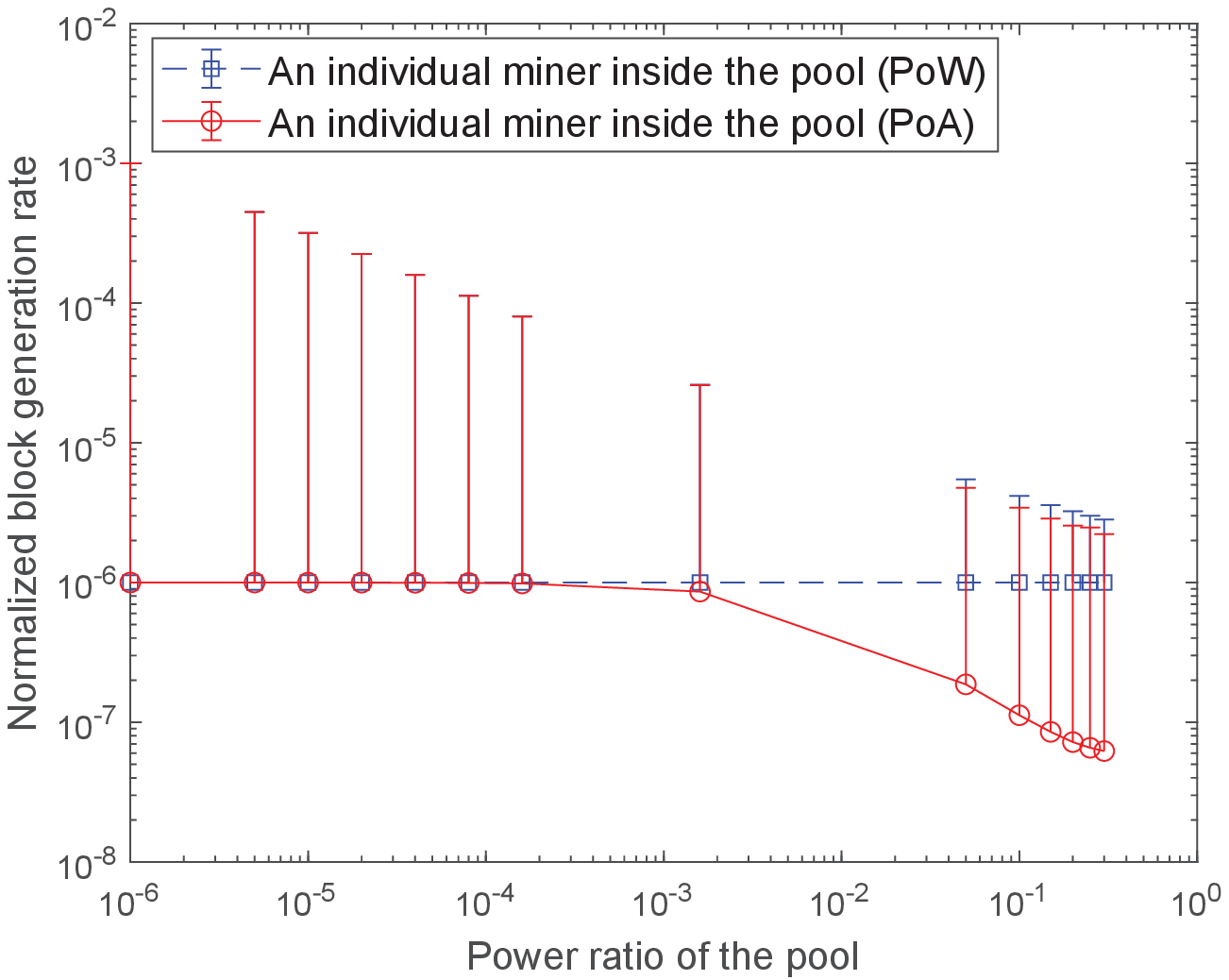}
       \caption{Comparsion of normalized block generation rate of an individual miner in the pool under PoW and PoA}
        \label{fig:simind}
\end{figure}

Fig. \ref{fig:simind} shows the normalized block generation rates of an individual miner inside the pool %and outside the pool (i.e., the solo miner)
under PoW and PoA respectively, where the y-axis is the block generation rate of an individual miner and the x-axis is the power ratio of the mining pool. First, under PoA,  the block generation rate of each individual miner inside the pool decreases as the pool's power ratio increases. This is due to a fact that as the increase of the pool's power ratio (i.e., the increase of the number of the miners in the pool), the increase of block generation rate of the pool is  negligible.  Second, under PoW, the block generation rate of each individual miner inside the pool is the same as that outside the pool, and keeps invariant over the range of power ratio; however, its variance is deceased by participating in the pool.

\section{Conclusion}
In this paper, we have proposed a novel consensus protocol, called PoA, to disincentivize the pooled mining. Following the native PoW protocol, the miners under PoA reap the rewards by solving the mathematical puzzles. In contrast to PoW, we introduced AoW to measure the effective mining period that the miner has spent solving the puzzles, and included the age ring in the block head as the input of hash puzzle to prove AoW. Regarding the protocol design, we proposed two different mining targets for the effective age ring and block respectively. In each mining round, the AoW will be increased by one if  the miner can solve the puzzle of age ring (i.e., the age ring is effective). The miner will be rewarded by either the reduced difficulty of block mining if its accumulated AoW is beyond a threshold, or the block reward if it generates a block. To facilitate the performance analysis, we use the CTMC to model the mining process of the mining pool and solo miners under PoA. Based on this CTMC, we show that the block generation rate of the pool under PoA can be reduced to 75$\%$ of that under PoW by a proper difficulty adjustment strategy. In addition, we not only approximated the distribution of block generation time of the mining pool, but also validated that the block generation time of all solo miners is exponentially distributed. Finally, we used the even-driven simulations to demonstrate the consistency between the numerical and simulation results.

\section*{Acknowledgement}
The authors would like to thank Dr. He Chen for his valuable discussions on the topic of age-of-information, which inspires the idea of this work.

\ifCLASSOPTIONcaptionsoff
\newpage
\fi

\bibliographystyle{IEEEtran}

\bibliography{refs}

\begin{thebibliography}{10}
\providecommand{\url}[1]{#1}
\csname url@samestyle\endcsname
\providecommand{\newblock}{\relax}
\providecommand{\bibinfo}[2]{#2}
\providecommand{\BIBentrySTDinterwordspacing}{\spaceskip=0pt\relax}
\providecommand{\BIBentryALTinterwordstretchfactor}{4}
\providecommand{\BIBentryALTinterwordspacing}{\spaceskip=\fontdimen2\font plus
\BIBentryALTinterwordstretchfactor\fontdimen3\font minus
  \fontdimen4\font\relax}
\providecommand{\BIBforeignlanguage}[2]{{%
\expandafter\ifx\csname l@#1\endcsname\relax
\typeout{** WARNING: IEEEtran.bst: No hyphenation pattern has been}%
\typeout{** loaded for the language `#1'. Using the pattern for}%
\typeout{** the default language instead.}%
\else
\language=\csname l@#1\endcsname
\fi
#2}}
\providecommand{\BIBdecl}{\relax}
\BIBdecl

\bibitem{nakamoto2008bitcoin}
S.~Nakamoto, ``Bitcoin: A peer-to-peer electronic cash system,''
  \emph{[Online]. Available: http://bitcoin.org}, 2008.

\bibitem{fanning2016blockchain}
K.~Fanning and D.~P. Centers, ``Blockchain and its coming impact on financial
  services,'' \emph{Journal of Corporate Accounting \& Finance}, vol.~27,
  no.~5, pp. 53--57, 2016.

\bibitem{dai2019blockchain}
H.-N. Dai, Z.~Zheng, and Y.~Zhang, ``Blockchain for internet of things: A
  survey,'' \emph{IEEE Internet of Things Journal}, vol.~6, no.~5, pp.
  8076--8094, 2019.

\bibitem{abeyratne2016blockchain}
S.~A. Abeyratne and R.~P. Monfared, ``Blockchain ready manufacturing supply
  chain using distributed ledger,'' \emph{International Journal of Research in
  Engineering and Technology}, vol.~5, no.~9, pp. 1--10, 2016.

\bibitem{natoli2019deconstructing}
C.~Natoli, J.~Yu, V.~Gramoli, and P.~Esteves-Verissimo, ``Deconstructing
  blockchains: A comprehensive survey on consensus, membership and structure,''
  \emph{arXiv preprint}, 2019, [Online]. Available:
  https://arxiv.org/abs/1908.08316.

\bibitem{BTCcom}
``Pool share from {BTC}.com,'' https://btc.com/stats/pool, Accessed March,
  2021.

\bibitem{eyal2014majority}
I.~Eyal and E.~G. Sirer, ``Majority is not enough: Bitcoin mining is
  vulnerable,'' in \emph{Proc. Int. Conf. Financial Cryptogr. Data
  Secur.}\hskip 1em plus 0.5em minus 0.4em\relax Springer, 2014, pp. 436--454.

\bibitem{chen2020age}
H.~Chen, Y.~Gu, and S.-C. Liew, ``Age-of-information dependent random access
  for massive iot networks,'' in \emph{Proc. IEEE INFOCOM 2020-IEEE Conference
  on Computer Communications Workshops (INFOCOM WKSHPS)}.\hskip 1em plus 0.5em
  minus 0.4em\relax IEEE, 2020, pp. 930--935.

\bibitem{kosta2017age}
A.~Kosta, N.~Pappas, and V.~Angelakis, ``Age of information: A new concept,
  metric, and tool,'' \emph{Foundations and Trends in Networking}, vol.~12,
  no.~3, pp. 162--259, 2017.

\bibitem{lewenberg2015inclusive}
Y.~Lewenberg, Y.~Sompolinsky, and A.~Zohar, ``Inclusive block chain
  protocols,'' in \emph{Proc. Int. Conf. Financial Cryptogr. Data Secur.}\hskip
  1em plus 0.5em minus 0.4em\relax Springer, 2015, pp. 528--547.

\bibitem{li2018scaling}
C.~Li, P.~Li, D.~Zhou, W.~Xu, F.~Long, and A.~Yao, ``Scaling nakamoto consensus
  to thousands of transactions per second,'' \emph{arXiv preprint
  arXiv:1805.03870}, 2018.

\bibitem{sompolinsky2015secure}
Y.~Sompolinsky and A.~Zohar, ``Secure high-rate transaction processing in
  bitcoin,'' in \emph{Proc. Int. Conf. Financial Cryptogr. Data Secur.}\hskip
  1em plus 0.5em minus 0.4em\relax Springer, 2015, pp. 507--527.

\bibitem{sompolinsky2016spectre}
Y.~Sompolinsky, Y.~Lewenberg, and A.~Zohar, ``Spectre: A fast and scalable
  cryptocurrency protocol.'' \emph{IACR Cryptol. ePrint Arch.}, vol. 2016, p.
  1159, 2016.

\bibitem{sompolinsky2018phantom}
Y.~Sompolinsky and A.~Zohar, ``Phantom,'' \emph{IACR Cryptology ePrint Archive,
  Report 2018/104}, 2018.

\bibitem{eyal2016bitcoin}
I.~Eyal, A.~E. Gencer, E.~G. Sirer, and R.~Van~Renesse, ``{Bitcoin-NG}: A
  scalable blockchain protocol,'' in \emph{Proc. 13th USENIX Symposium on
  Networked Systems Design and Implementation (NSDI)}, 2016, pp. 45--59.

\bibitem{rizun2016subchains}
P.~R. Rizun, ``Subchains: A technique to scale bitcoin and improve the user
  experience,'' \emph{Ledger}, vol.~1, pp. 38--52, 2016.

\bibitem{TierNolan}
TierNolan, ``Decoupling transactions and pow,''
  https://bitcointalk.org/index.php?topic=179598.0, Accessed March, 2021.

\bibitem{bagaria2019prism}
V.~Bagaria, S.~Kannan, D.~Tse, G.~Fanti, and P.~Viswanath, ``Prism:
  Deconstructing the blockchain to approach physical limits,'' in \emph{Proc.
  the 2019 ACM SIGSAC Conf. Comput. Commun. Secur. (CCS)}, 2019, pp. 585--602.

\bibitem{pass2017fruitchains}
R.~Pass and E.~Shi, ``Fruitchains: A fair blockchain,'' in \emph{Proc. the ACM
  Symposium on Principles of Distributed Computing}, 2017, pp. 315--324.

\bibitem{P2pool}
``{P2Pool}: Decentralized bitcoin mining pool,'' http://p2pool:org/, Accessed
  March, 2021.

\bibitem{luu2017smartpool}
L.~Luu, Y.~Velner, J.~Teutsch, and P.~Saxena, ``{SmartPool}: Practical
  decentralized pooled mining,'' in \emph{Proc. 26th USENIX Security
  Symposium}, 2017, pp. 1409--1426.

\bibitem{xue2018proof}
T.~Xue, Y.~Yuan, Z.~Ahmed, K.~Moniz, G.~Cao, and C.~Wang, ``Proof of
  contribution: A modification of proof of work to increase mining
  efficiency,'' in \emph{Proc. 2018 IEEE 42nd Annual Computer Software and
  Applications Conference (COMPSAC)}, vol.~1.\hskip 1em plus 0.5em minus
  0.4em\relax IEEE, 2018, pp. 636--644.

\bibitem{nayak2016stubborn}
K.~Nayak, S.~Kumar, A.~Miller, and E.~Shi, ``Stubborn mining: Generalizing
  selfish mining and combining with an eclipse attack,'' in \emph{Proc. IEEE
  Eur. Symp. Secur. Privacy (EuroS\&P)}, 2016, pp. 305--320.

\bibitem{carlsten2016impact}
M.~Carlsten, ``The impact of transaction fees on bitcoin mining strategies,''
  Master's thesis, Princeton University, 2016.

\bibitem{sapirshtein2016optimal}
A.~Sapirshtein, Y.~Sompolinsky, and A.~Zohar, ``Optimal selfish mining
  strategies in bitcoin,'' in \emph{Proc. Int. Conf. Financial Cryptogr. Data
  Secur.}, 2016, pp. 515--532.

\bibitem{sompolinsky2016bitcoin}
Y.~Sompolinsky and A.~Zohar, ``Bitcoin's security model revisited,''
  \emph{arXiv preprint arXiv:1605.09193}, pp. 1--26, 2016.

\bibitem{gervais2016security}
A.~Gervais, G.~O. Karame, K.~W{\"u}st, V.~Glykantzis, H.~Ritzdorf, and
  S.~Capkun, ``On the security and performance of proof of work blockchains,''
  in \emph{Proc. the 2016 ACM SIGSAC Conf. Comput. Commun. Secur. (CCS)}, 2016,
  pp. 3--16.

\bibitem{wang2019survey}
W.~Wang, D.~T. Hoang, P.~Hu, Z.~Xiong, D.~Niyato, P.~Wang, Y.~Wen, and D.~I.
  Kim, ``A survey on consensus mechanisms and mining strategy management in
  blockchain networks,'' \emph{IEEE Access}, vol.~7, pp. 22\,328--22\,370,
  2019.

\bibitem{rosenfeld2011analysis}
M.~Rosenfeld, ``Analysis of bitcoin pooled mining reward systems,'' \emph{arXiv
  preprint arXiv:1112.4980}, 2011.

\bibitem{anderson2012continuous}
W.~J. Anderson, \emph{Continuous-time Markov chains: An applications-oriented
  approach}.\hskip 1em plus 0.5em minus 0.4em\relax Springer Science \&
  Business Media, 2012.

\end{thebibliography}

\end{document}